\documentclass[rnote,traditabstract]{aa} % for the research notes
\usepackage{graphicx}
\usepackage{txfonts}
\begin{document}
   \title{Confidence Level Estimator for cosmological models}

   \author{G. Sironi}

   \institute{Dipartimento di Fisica G.Occhialini, University of Milano Bicocca, Piazza della Scienza 3, 20126 Milano\\
              \email{giorgio.sironi@unimib.it}}

\date{Received September xx, 2010; accepted xx yy, 201x}

% \abstract{}{}{}{}{}
% 5 {} token are mandatory

 \abstract{
Models of the Universe like the Concordance Model today used to
interpret cosmological observations give expectation values for
many cosmological observables so accurate that frequently peoples
speak of Precision Cosmology. The quoted accuracies however do not
include the effects of the priors used in optimizing the Model nor
allow to evaluate the confidence one can attach to the Model. We
suggest an estimator of the Confidence Level for Models and the
accuracies of the expectation values of the Model observables.}

  \keywords{cosmology: cosmological parameters -- cosmology: observations --
  cosmology: miscellaneous}

   \maketitle
%
%________________________________________________________________

\section{Introduction}
After the discovery of the Cosmic Microwave Background (CMB) in
1964 (\cite{penw64}), cosmological observation went through a very
rapid expansion. Measurements confirmed to very high degree of
accuracy the expected first order (planckian spectrum, isotropic
distribution, absence of polarization) and second order (angular
power spectra of tiny temperature an\-iso\-tro\-pies and even
smaller level of residual polarization) features of the CMB
(\cite{cobe,wmap}). Complementary information then arrived from
no-CMB channels of information, namely: i)peculiarities in the
rotation curves of galaxies and clusters of galaxies, (which
suggest the existence of "invisible" or "dark" matter),
ii)discovery of BAO (Barion Acoustic Oscillations) in the spatial
distribution of the luminous matter, similar to the oscillations
observed in the CMB angular power spectra, iii)Supernovae Ia used
to get the distance of objects at very large Z, iv)N-body
simulations of the formation of matter condensations from
isotropic distribution of matter particles (\cite{nocmb}).
\par The whole set of CMB and no-CMB information, besides confirming the general layout
of the Standard Big Bang Model (\cite{peebles}), suggests that the
Universe is probably suffering re-acceleration and its geometry
substantially flat (euclidean). To close the Universe it was
therefore assumed the existence, beside barionic and dark matter,
of a third component of the matter-energy mixture, Dark Energy.
Moreover besides the Hubble constant $H_o$, the ratio between the
primordial abundances of helium and hydrogen and the Universe
average matter density $\Omega$ (in units of the critical density
$\rho_c = 3 H_o^2/(8 \pi G)$), new observables were introduced:
the abundances (still in unit of critical density) of barionic
matter $\Omega_b$, dark matter $\Omega_{dm}$ and dark energy
$\Omega_\Lambda$, the optical thickness $\tau$ of the Universe at
reionization (when at $10 \leq Z_{ion} \leq  1300$ stable matter
condensation emerged from the primordial uniform matter
distribution), the amplitude and spectral index ( $A_s$ and $n_s$)
of the fluctuations (seed for the birth of the above
condensations), and $\sigma_8$ an indicator of the fluctuations of
the galactic matter distribution.
\par To reconcile these new data and
the old Big Bang model the Concordance Model or $\Lambda$-CDM
Model gradually emerged (\cite{concord,kova}). It has six
independent parameters, representative of six observable chosen
among the observables listed above, usually $H_o$, $\Omega_b$,
$\Omega_{dm}$, $\tau$, $A_s$ and $n_s$, from which other
observable (e.g age of the Universe $T_{univ}\simeq 1/H_o$,
critical density $\rho_c$ of matter-energy, Dark Energy density,
$\Omega_\Lambda$ (in unit of $\rho_c$), reionization red shift
$Z_{ion}$ and $\sigma_8$) can be derived.
\par The observables-parameters and the observational data are
entangled: different sets of data are produced by different
parameter combinations. Disentangling the effects of the various
parameters and getting the parameters values is extremely hard. A
way out is obtained fitting the Model to the full set of CMB and
no-CMB cosmological data and getting the parameter values by best
fit methods. This is done by Montecarlo methods using Markov
chains to implement the stochastic procedure which drives
Montecarlo calculations (\cite{montecarlo}) with additional {\it a
priori} assumptions, e.g. the Universe geometry ($\Omega =
\Omega_b + \Omega_{dm} + \Omega_\Lambda =1$ if the Universe is
flat) and$/$or intervals of allowed variability for the
parameters. The parameter values which give the best fit are
interpreted as expectation values of the associated observables.
\par This procedure, which forces the model toward preferred solutions, is now well
established. Frequently improved expectation values are published.
They are obtained adding new observational data to the preexisting
data base (e.g.\cite{fiveyear,brow09,sevenyear}). Recently
published values (\cite{sevenyear}) of free and derived observable
- parameters of the Concordance Model are shown in Table \ref{T1}
and Table \ref{T2}. The accuracy of the expectation values is so
high that observers sometimes  speak of Precision Cosmology (e.g.
\cite{precision,bridle}).
\par In principle all the observables are measurable.
Agreement between measured and expected values is a proof that the
model described by the free parameters is a good one.
Unfortunately for five of the six free observables - parameters of
the Concordance Model measurements are very poor or not yet
available. And no significant improvement is expected in the near
future. For these observable in literature there are only large
intervals inside which the measured values are expected. The same
intervals are usually assumed as the maximum variability range of
the parameters used in Monte Carlo studies of the Concordance
Model (\cite{brow09}). They are shown in Table \ref{T1}. The only
exception is the Hubble constant for which almost direct
measurements are now very accurate (\cite{hubble}).
\section{Expected and measured values}
\par Measured values $M^i$ and their uncertainties $\delta_{M^i}$
are extracted using classical statistical methods (see for
instance \cite{bevi}) from sets of repeated measurements of a
parameter $i$ or from combinations of measured quantities directly
linked to $i$. The distribution of the measured values around
their average $\overline{M^i}$ is usually gaussian.
\par Expected values $E^i_v$ are the values of the model
parameters which gives the best fit of the Model to the full set
of observations assuming {\it a priori} conditions or {\it
priors}. The uncertainties $\delta_{E_v}$ of $E_v$ are the widths
of the distributions (not always gaussian) which encompass 68 $\%$
of the parameter values obtained repeating the calculation with a
random initial choice inside {\it a priori} chosen intervals of
free parameter variabilities.

\begin{table*}
\center{ \caption{$\Lambda CDM$-Concordance Model: Expectation
values, Measured Values and Confidence Level of Model Parameters -
(adapted from \cite{sevenyear,brow09})} \label{T1} \centering
\begin{tabular}{llccc}
&&&&\\ \hline &&&&\\ Parameter &&$<E>$& M & $C_l$ \\ && Expec.
value &
Meas. value&Conf. Level \\ &&&&\\ \hline &&&&\\
Hubble Constant (km/sec Mpc)&$H_o$ & $70.4^{+1.3}_{-1.4}$ &
$74.2\pm3.6$& 0.38 \\ &&&&\\
Barionic Matter Density & $\Omega_b$ & $0.0456 \pm 0.0016$ &
$0.005 - 0.1$& $ < 10^{-2}$ \\ &&&&\\
Dark Matter density & $\Omega_{dm}$ & $0.227 \pm 0.0014$ & $0.006
- 1$&$ < 10^{-2}$ \\ &&&&\\
Optical thickness at reionization & $\tau$ & $0.087 \pm 0.0014$ &
$0.01 - 0.80$&$ < 10^{-2}$ \\ &&&&\\
Scalar fluctuations Amplitude & $A_s$ & $(2.441
^{+0.088}_{-0.092}~10^{-9})$ &?&$?$ \\ &&&&\\
Scalar Spectral index
& $n_s$ & $0.963 \pm 0.012$ & $0.5 -
1.5$&$ < 2~10^{-2}$ \\
&&&&\\
\hline &&&&\\
\end{tabular}
}
\end{table*}

\begin{table*}
\center{ \caption{$\Lambda CDM$-Concordance Model: Expectation
values of Derived Parameters - (adapted from
\cite{sevenyear,brow09})} \label{T2} \centering
\begin{tabular}{lclcc}
&&&&\\ \hline &&&&\\
Parameter &&& & Expected value \\ &&&&\\ \hline
&&&&\\
Dark Energy Density && $\Omega_\Lambda$ &&
$0.728^{+0.015}_{-0.016}$ \\ &&&&\\
Reionization Red Shift && $Z_{ion}$ && $10.4 \pm 1.2$ \\
&&&&\\
galactic fluctuations amplitude && $\sigma_8$ && $0.809 \pm 0.024$
\\ &&&&\\
Universe Age (years) && $t_o$ && $(13.75 \pm 0.11)~10^9$ \\ &&&&\\
&&&& \\
\hline &&&&\\
\end{tabular}
}
\end{table*}

In doing it, sometimes unintentionally, a transition from
classical to bayesian statistics occurred (\cite{bayes}). Bayesian
probabilities, no longer precisely defined as frequencies or
ratios of measured quantities, give the (0-1) confidence level we
associate to an event occurrence and are obtained by {\it a
priori} assumptions. The Bayes Theorem allows to evaluate the
effects new observations have on preexisting knowledge:
\begin{equation}
P_E(H) ~=~ [P(H)/P(E)] P_H(E)
\end{equation}
\par\noindent where
\par i)H is a set  of measured values of a parameter and $P(H)$
its probability distribution, derived for instance from previous
observations, assumed {\it ab initio}, i.e. before new
observations are made;
\par ii)E is a new set of measurements of the same parameter
and $P(E)$ its distribution assumed  {\it a priori}, or {\it
prior}, without preexisting information;
\par iii) $P_H(E)$ is the conditional or "direct" probability of getting E given
H;
\par iv) $P_E(H)$ is the {\it a posteriori} conditional probability of
getting H given E, or "inverse" probability.
\par\noindent The ratio $0 \leq [P(H)/P(E)] \leq 1$ is the Confidence
level $C_l(E)$ we can associate to the prior used to improve our
knowledge.
\par For each parameter $i$ of a multi parameter model we can
write
\begin{eqnarray}
P^i(H) &=& P_{me}^i(X^i_{me}, \sigma^i_{me}) \\
P^i(E) &=& P_{ex}^i(X^i_{ex}, \sigma^i_{ex})
\end{eqnarray}
\par\noindent where $P_{me}^i(X^i_{me}, \sigma^i_{me})$ and
$P_{ex}^i(X^i_{ex}, \sigma^i_{ex})$ are the probabilities
distributions of the measured (pedix $_{me}$) and expected (pedix
$_{ex}$) values $X^i$, with dispersion $\sigma^i$. Assuming
gaussian distributions the Confidence level of the expected value
of parameter $i$ can be written:
\begin{eqnarray}
C_l^i &=& [P^i(H)/P^i(E)]  \nonumber \\ \nonumber \\&=&
P_{me}^i(X^i_{me}, \sigma^i_{me})/P_{ex}^i(X^i_{ex},
\sigma^i_{ex})\\ \nonumber \\
&=&\frac{\sigma_{ex}^i}{\sigma_{me}^i}~exp\{-[(\Delta
X^i_{me}/\sigma^i_{me})^2 - (\Delta X^i_{ex}/\sigma^i_{ex})^2]/2\}
\nonumber
\end{eqnarray}
\par\noindent where $\Delta X^i = |X^i - \overline{X^i}|$.
\par For a model not too far from the
real world $$<E^i>~\longrightarrow ~M^i, ~~~~\Rightarrow~~~~\Delta
X^i_{me} \simeq \Delta X^i_{ex} \simeq \Delta X^i < \sigma^i$$
\par\noindent therefore
\begin{equation}
C_l^i = \frac{\sigma_{ex}^i}{\sigma_{me}^i}~\big[1~-~\frac{\Delta
X^i}{2}~\frac{(\sigma_{me}^i)^2
-(\sigma_{ex}^i)^2}{(\sigma_{me}^i)^2~(\sigma_{ex}^i)^2}\big]
\rightarrow  \sigma^i_{ex}/\sigma^i_{me}
\end{equation}
\par\noindent for parameter {\it i} and for the whole model
\begin{equation}
C_l = \frac{1}{6}~\sum_{i=1}^6 C_l^i
\end{equation}
\par The resulting Confidence Levels of the expected values of the
free parameters of the Concordance Model are shown in the last
column of Table \ref{T1}. For the whole Model
$$ C_l ~\sim 0.4$$
\par\noindent a value which leaves room for other models to be considered,
obtained for instance relaxing priors, e.g. $\Omega_b + \Omega_c +
\Omega_\Lambda =1$.
\section{Discussion} The numerical values of
$C^i_l$ and $C_l$ presented above have been obtained for the
$\Lambda$-CDM Concordance Model of the Universe. The difference
between measured values and expectation values of the observable -
parameters however hold for other Models. For each model we can
evaluate expected value and Confidence Level $C^i_l$ of each
observable and $C_l$ for the whole Model. The uncertainties of the
calculated Expectation values can be assumed as uncertainties on
the observable real values only when the model {\it priors} are
such that $\sigma_{me} \simeq \sigma_{ex}$.
\par If not, systematic uncertainties introduced
by the {\it priors} have to be added to the uncertainties of the
observable.
\par A way to validate priors and model is therefore repeating direct
(or almost direct) measurements of the parameters until their
accuracies becomes comparable to the accuracy of the expected
values.
\par\noindent Until  $C_l < 1$ models based on different priors
cannot be excluded.
\section{Conclusion} The use of Montecarlo
methods and Bayesian Statistics to analyze the enormous quantity
of data of cosmological interest which are continously poured by
ground and space observations is almost unavoidable. However
Montecarlo and Bayesian Methods are based on {\it a priori}
assumption whose statistical weight should be added, but rarely is
added, to the the quoted accuracies of the parameter expected
values.
\par Peoples who currently use these methods are aware of that
and warning have been put forward (e.g. \cite{bridle}).
Unfortunately general public and professionals not involved in
cosmological observations may be unaware of it, misinterpret the
results of model simulations and attribute weights above their
real values to models. Forgetting it may stop or reduce support to
studies of other models based on different {\it priors} still
possible and not yet excluded by observation.
\par Similar situations occur in other fields of pure and applied
research (e.g. unification of fundamental forces, string theories,
elementary particle models, models of climate evolution and so
on).  To avoid misunderstandings and preserve possibilities of
pursuing alternatives lines of research, this point of view must
be transmitted to the bodies who support research activities and
to the general public.
\begin{acknowledgements} Observations of the CMB spectrum have been
carried out by the Milano Radio Group with the support of MIUR
(Italian Ministry of University and Research), CNR (Italian
National Council of Research), PNRA (Italian Program for Antarctic
Research).
\end{acknowledgements}


\begin{thebibliography}{Dillo 83}

% \harvarditem{Name}{Year}{label}
% Text of bibliographic item
\bibitem[Bevington and Robinson 1992]{bevi} Bevington, P.R. and
Robinson, K.D. Data Reduction and Error Analysis for the Physical
Sciences, Mc Graw-Hill, London 1992
\bibitem[Bridle et al. 2003]{bridle} Bridle, S.L., Lahav, O.,
Ostriker, J.P. (2003) Sci 299, 1532
\bibitem[Brown et al. 2009]{brow09} Brown, M.L. et al. (2009) ApJ 705,
978
\bibitem[COBE 2000]{cobe}http://lambda.gsfc.nasa.gov/product/cobe/
and references therein
\bibitem[Komatsu et al. 2009]{fiveyear} Komatsu E. et al. ApJS
(2009) 180, 330
\bibitem[Komatsu et al. 2010]{sevenyear} Komatsu, E. et al,.
arXiv:1001.4538v1
\bibitem[Kowalski et al. 2008]{kova} Kowalski M. et al. (2008) ApJ 686,
749
\bibitem[Landau and Binder 2009]{montecarlo} Landau D.P. and
Binder K. A guide to Monte-Carlo Simulations in Statistical
Physics, Cambridge University Press, Cambridge UK 2009
\bibitem[LSS 2010]{nocmb} http://lambda.gsfc.nasa.gov/product/lss/
and references therein
\bibitem[Ostriker and Steinhardt 1995]{concord} Ostriker J.P. and
Steihardt P.J. arXiv:astro-ph/9505066v1
\bibitem[Peebles 1993]{peebles} Peebles P.J.E. Principles of
Physical Cosmology, Princeton University Press, Princeton USA 1993
\bibitem[Penzias and Wilson 1964]{penw64} Penzias, A.A., Wilson, R.W., ApJ (1965), 142, 419
\bibitem[Primack 2005]{precision} Primack J.R. New Astr. rev. (2005) 49,
25
\bibitem[Riess et al. 2009]{hubble} Riess, A.G. et al. (2009) ApJ699,
539
\bibitem[Stanford 2003]{bayes}
http://plato.stanford.edu/entries/bayes-theorem/ and references
therein
\bibitem[WMAP 2010]{wmap} http://wmap.gsfc.nasa.gov/ and references therein



\end{thebibliography}
\end{document}